\documentclass{article}
\usepackage[T1]{fontenc} 
\usepackage[utf8]{inputenc} 
\usepackage{ismir,amsmath,cite,url}
\usepackage{graphicx}
\usepackage{breqn}
\usepackage{color}
\usepackage{comment}
\usepackage{paralist}

\newcommand{\jordan}[1]{\textcolor{black}{#1}}

\usepackage{lineno}

\title{The Freesound Loop Dataset and Annotation Tool}

\multauthor
{Ant\'{o}nio Ramires$^1$ \hspace{1cm} Frederic Font$^1$ \hspace{1cm} Dmitry Bogdanov$^1$ \hspace{1cm} Jordan B. L. Smith$^2$} { \bfseries{Yi-Hsuan Yang$^3$ \hspace{1cm} Joann Ching$^3$ \hspace{1cm} Bo-Yu Chen$^3$ \hspace{1cm}} Yueh-Kao Wu$^3$\\ \bfseries{Hsu Wei-Han$^3$ \hspace{1cm} Xavier Serra$^1$}\\
$^1$ Music Technology Group, Universitat Pompeu Fabra, Barcelona, Spain\\
$^2$ TikTok, Lodon, United Kingdom\\
$^3$ Research Center for IT Innovation, Academia Sinica, Taipei, Taiwan\\
\tt\small antonio.ramires@upf.edu
}
\def\authorname{Ant\'{o}nio Ramires, Frederic Font, Dmitry Bogdanov, Jordan B. L. Smith, Yi-Hsuan Yang, Joann Ching, Bo-Yu Chen, Yueh-Kao, Hsu Wei-Han, Xavier Serra}

\begin{document}

\maketitle
\begin{abstract}

Music loops are essential ingredients in electronic music production, and there is a high demand for pre-recorded loops in a variety of styles. Several commercial and community databases have been created to meet this demand, but most are not suitable for research due to their strict licensing. We present the Freesound Loop Dataset (FSLD), a new large-scale dataset of music loops annotated by experts. The loops originate from Freesound, a community database of audio recordings released under Creative Commons licenses, so the audio in our dataset may be redistributed. The annotations include instrument, tempo, meter, key and genre tags. We describe the methodology used to assemble and annotate the data, and report on the distribution of tags in the data and inter-annotator agreement. We also present to the community an online loop annotator tool that we developed. To illustrate the usefulness of FSLD, we present short case studies on using it to estimate tempo and key, generate music tracks, and evaluate a loop separation algorithm. We anticipate that the community will find yet more uses for the data, in applications from automatic loop characterisation to algorithmic composition.

\end{abstract}

\section{Introduction}\label{sec:introduction}
Repurposing audio material to create new music\jordan{---}also known as \emph{sampling}\jordan{---}was a foundation of electronic music and is a fundamental component of this practice. Loops are audio excerpts, usually of short duration, that can be played repeatedly in a seamless manner \cite{LoopsasGenreResources}. These loops can serve as the basis for songs, which music makers can combine, cut and rearrange, and have been extensively used in Electronic Dance Music (EDM) tracks \cite{butler2006unlocking}. 

Audio loops have been made available for amateur and professional music makers since the early ages of electronic music. 
Currently, large-scale databases of audio offer huge collections of audio material for users to work with. Some databases, like Freesound\footnote{\url{https://freesound.org/}} and Looperman\footnote{\url{https://www.looperman.com/}}, are community-oriented: people upload their sounds so that other users can employ them \jordan{in} their works. More commonly, these collections are commercially oriented: loops are available to paying costumers, either through a 
subscription service (\emph{e.g.} Sounds.com,\footnote{\url{https://sounds.com/}} Splice\footnote{\url{https://splice.com/}}) or by allowing customers to buy packs of loops (\emph{e.g.} Loopmasters,\footnote{\url{https://www.loopmasters.com/}} and Prime Loops\footnote{\url{https://primeloops.com/}}).

Despite the number of loops available on these databases, the technologies used to analyse and navigate these databases still rely on human annotations and human content curation to, for instance, group sounds into packs for specific genres or styles. Loops are being manually annotated with information like instrument, tonality (key), tempo (bpm) and music genre. This is a time-consuming task which is often unfeasible, which results in badly annotated databases and poor user experience when browsing them. In the field of Music Information Retrieval (MIR), a substantial effort \jordan{has been put into} automatically identifying the aforementioned characteristics for musical pieces. However, loops are inherently different from music pieces (\emph{i.e.} with reduced instrumentation and short length). Therefore, existing MIR algorithms need to be tested and (possibly) adapted to work successfully in this scenario. Furthermore, new MIR tasks are emerging with the study of music loops including
loop retrieval~\cite{drumRetrievalSpokenQuery},
loop detection~\cite{lopez-serrano2016},
loop discovery~\cite{lopez-serrano2017} and extraction~\cite{loopsep},
loop recommendation~\cite{chen20ismir},
exploration of large loop databases~\cite{08icme_loop_explore},
and automatic loop generation~\cite{18chiloopmaker}.

In this paper, we present FSLD, an open dataset with 9,455 music loops to support reproducible research in MIR. FSLD contains production-ready loops from Freesound which are distributed under Creative Commons licenses and can, therefore, be freely shared among the research community and industry. Part of the dataset has been manually annotated with information about rhythm, tonality, instrumentation and genre, in a similar way as commercially available loop collections are annotated. The annotation service is made public\footnote{\label{annotator}\url{http://mtg.upf.edu/fslannotator}} so that the community can work on enlarging the annotations of this collection. We expect this dataset to have an impact on the research community as it supports further research into several timely research topics which are also of great interest to the industry. 

\begin{figure*}[!t]
  \includegraphics[width=.95\textwidth]{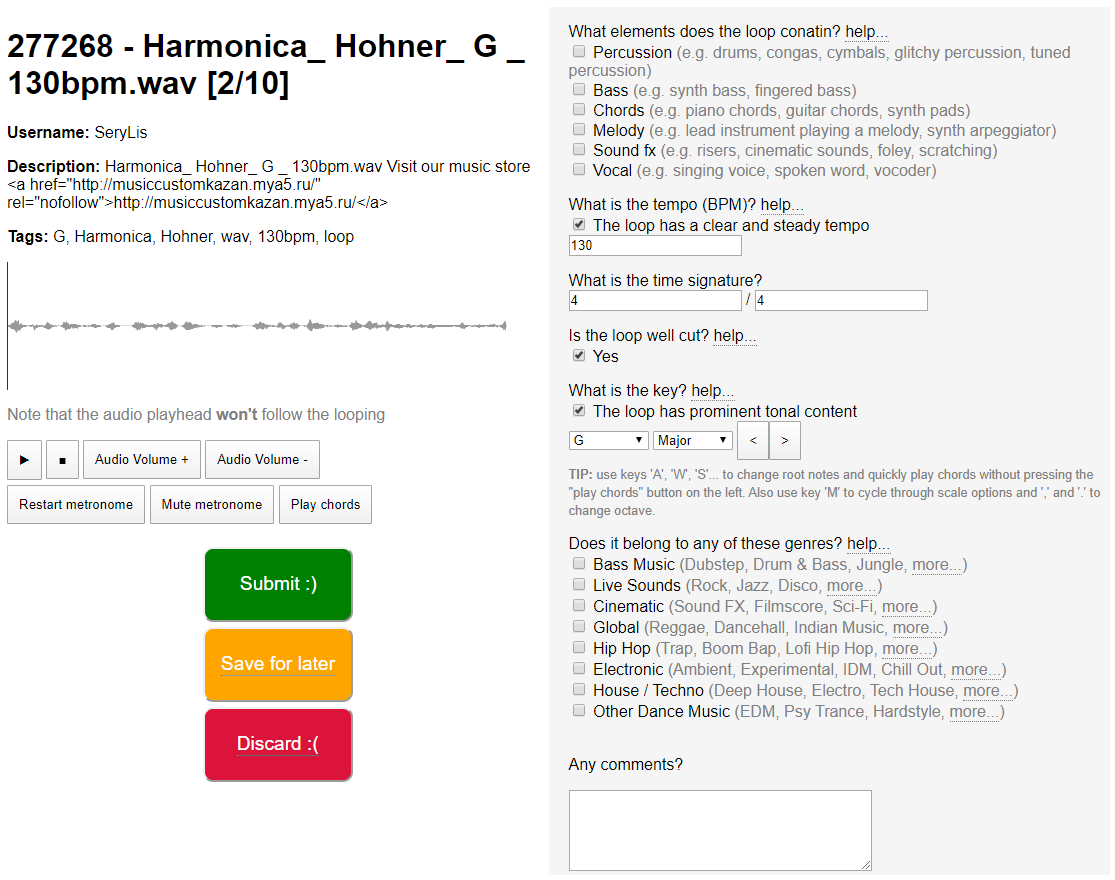}
  \caption[UI]{The user interface provided to the annotators. }
  \label{img:annotationtool}
\end{figure*}

The rest of the paper is structured as follows. In Sec. \ref{sec:relatedwork}, we present some of the datasets used in the literature for loop analysis. Sec. \ref{sec:datasetcreation} details how the proposed dataset was collected and annotated. In Sec. \ref{sec:datasetanalysis}, general statistics of the dataset are given. In Sec. \ref{pot_app1} and \ref{pot_app2}, we present some potential applications and provide a benchmark of the dataset using some classic MIR tasks. Finally, in Sec. \ref{sec:conclusion}, we conclude and suggest future work directions.



\section{Related Work}\label{sec:relatedwork}

Early work on the retrieval of loops focused on tempo extraction and transcription from drum loops \cite{automaticTranscription, experimentalComparison}. Gouyon et al. compared several tempo induction algorithms proposed in the ISMIR 2004 competition \cite{experimentalComparison}. The loop dataset used in this work has been commonly used for evaluating tempo estimation algorithms and is divided into three subsets. One of these comprises two thousand audio loops (with tempo annotations) from Sound Effects Library.\footnote{\url{http://www.sound-effects-library.com/}} These audio loops are not free and a license needs to be obtained to use them for research.

Automatic transcription of drum loops focuses on identifying when the different percussion instruments occur in a loop. Gillet \jordan{and Richard} used a collection of 315 drum loops for evaluating their system and provided ``a compressed version of a few drum loops''~\cite{automaticTranscription}. The URL to the webpage the authors provide is broken and, presumably, the lower quality versions of the loops do not represent commercial-quality content. The authors also use this dataset for automatically retrieving drum loops from spoken queries \cite{drumRetrievalSpokenQuery}. This database was later used by Bello et al. for automatic rhythm modification and analysis of drum loops \cite{modificationRhythmLoops,analysisRhythmLoops}.

The work from Gómez-Marín et al. explores rhythmic similarity measures for audio loops \cite{gomezMarinPadSad}. The authors validate the proposed metric using 9 drum break loops from Rhythm Lab.\footnote{\url{https://rhythm-lab.com/}} The authors do not specify which are the drum loops used.

Font et al. presented a dataset of audio loops from Freesound \cite{FS} in their work on tempo estimation and a confidence measure for audio loops \cite{fontTempo}. 
The authors use two commercial datasets, loops bundled with music production software Apple's Logic Pro\footnote{\url{https://www.apple.com/logic-pro/}} and Acoustica's Mixcraft,\footnote{\url{https://acoustica.com/mixcraft}} and two community datasets. The first one is a private collection of loops downloaded from Looperman, which was previously used for research in \cite{roma_gerard_2015_3674147}. Looperman does not allow the re-distribution of loops ``as is'', and considers as misuse the automatic download of their loops.\footnote{\url{https://www.looperman.com/help/terms}} A collection of 4000 loops from Freesound, obtained by searching Freesound for sounds with the queries ``loop'' and ``bpm'' is also proposed. The sounds' filenames, tags and textual descriptions are parsed to identify tempo annotations provided by the users. However, these annotations are not always accurate, and, to enable further work on audio loops, more information besides the tempo is desired.

In short, existing academic work which employs loops resorts to commercial samples as the source of data and open datasets do not have complete and reliable annotations.
\jordan{This makes it difficult to reproduce existing research.}
To promote open and accessible research on audio loops, we propose a free and distributable database of loops from Freesound, 
which provides production-ready sounds with high-quality annotations.


    
\section{Dataset Creation}\label{sec:datasetcreation}

In this section the process we have followed to create the dataset is described. We show how we collected the loops to annotate, how they were pre-analysed for a faster annotation procedure and explain what was annotated and how the annotation tool was implemented. Finally, we present how the dataset is distributed and organised. 

\subsection{Loop Selection}

To select an initial pool of candidate loops, we followed the same methodology as in~\cite{fontTempo}: i.e., we retrieved sounds with both ``loop'' and ``bpm'' keywords on Freesound, resulting in 9,490 sounds.
Using the Freesound API, it was straightforward to obtain these loops and their metadata---title, tags, textual description, and author’s username.


\subsection{Loop Annotation}

We want the loops in our dataset to be annotated in a way which is similar to commercially available loops. This way, we make sure that the loop characterisation is compatible with industry standards. For this, we decided to annotate the loops' instrumentation, tempo, time signature, key and genre, as described below. The annotation was performed by 8 MIR researchers and students, with knowledge of electronic music production. To make the annotation procedure as efficient as possible, we created a web application for the annotators with several tools at their disposal, which can be seen in Fig. \ref{img:annotationtool}. This application was developed using Flask,\footnote{\url{https://flask.palletsprojects.com/}} a web framework for Python. 

This interface provides fields for the annotators to fill in the desired information, which will be described in the following sections. The instructions are provided on tooltips for quick access by annotators.

\subsubsection{Instrumentation}

Instead of annotating
instruments in a traditional way, which would not be straightforward in heavily processed audio or more experimental loops, we chose to annotate general \textit{roles} which can be useful for both music makers and automatic generation of music. We asked annotators to tick all the roles that apply to each loop. Usually, specific instruments could be easily assigned to a specific role. We present the roles along with some examples in Table \ref{tab:inst}.

\begin{table}[h]
\small
\begin{tabular}{l|l}
\textbf{Role} & \textbf{Example Instruments} \\
\hline
Percussion & Drums, glitches, tuned percussion \\
Bass & Synth bass, fingered bass\\
Chords & Piano chords, guitar chords, synth pads\\
Melody & Instrument playing a melody, arpeggiator\\
Sound FX & Risers, cinematic sounds, foley, scratching \\
Vocal & Singing voice, spoken word, vocoder
\end{tabular}
\caption{Instrumentation roles and the examples provided for each category.} \label{tab:inst}
\end{table}

\subsubsection{Rhythmic Characteristics}
We asked for annotations on three rhythmic aspects:

\textbf{Tempo} provides an easy measure of rhythmic compatibility and is the most common information provided in commercial loop databases. We ask annotators if the loop has a clear and steady tempo, to identify loops with constant tempo and clear beat (BPM value and steady tempo), with changing tempo (BPM value of the initial tempo and no steady tempo), and loops with no clear beat but where the tempo can be inferred (BPM value and no steady tempo). 

\textbf{Meter} is not a feature we see annotated as often as BPM, which might be due to the common use of 4/4 meter in electronic music. This feature is relevant to annotate, for calculating the number of bars in a loop, from its meter, tempo and duration. 

Finally, as sometimes the length of the audio file is not the length of the loop, we also annotate if it is \textbf{well-cut}. If there is some silence at the beginning or the end of the file or if there is a ``tail'' (\emph{e.g.} a decay of a reverb effect) when the audio is exported, it might not loop correctly just by staring the loop again when it finishes playing. 

\subsubsection{Tonal Characteristics}
We annotate if the loop has prominent tonal content and, if so, to indicate a root key and mode that matches the tonal content of the loop (\emph{i.e.}, root note from a chromatic scale and \emph{Major}/\emph{Minor} mode). We explained ``prominent tonal content'' as whether it is easy to sing along to the loop or to find a meaningful root note for the loop. For root key annotations, we asked to choose a note from a dropdown with 12 notes, or ``Unknown'' in case the key could not be found. For annotating mode, the annotators had the choice of ``Major'' or ``Minor'' if the loop sounded good with one of these modes; ``None'' if the loop could not be clearly assigned to either ``Major'' or ``Minor'' (e.g. loop contains a single note); or ``Unknown'' for other cases. 

\subsubsection{Genre}\label{subsubsec:genre}
We annotate genre in non-exclusive categories, where each is assigned to a loop if it can be used to make music in that genre. 
To create a taxonomy which would be similar to commercially available ones, we merged the taxonomies of Sounds.com and Splice. These were chosen as they provided several examples for each genre and had similar parent categories. We present the taxonomy in Table \ref{tab:genres}.

\begin{table}[h]
\small
\begin{tabular}{l|l}
\textbf{Genre} & \textbf{Examples} \\
\hline
Bass Music & Dubstep, Drum and Bass, Jungle \\
Live Sounds & Rock, Jazz, Disco \\
Cinematic & Sound FX, Filmscore, Sci-Fi \\
Global & Reggae, Dancehall, Indian Music \\
Hip Hop & Trap, Boom Bap, Lofi Hip Hop \\
Electronic & Ambient, IDM, Chill Out \\
House / Techno & Deep House, Electro, Tech House \\
Other Dance Music & EDM, Psy Trance, Hardstyle \\
\end{tabular}
\caption{Taxonomy of genres used for the annotation and examples for each category.}
\label{tab:genres}
\end{table}

\subsubsection{Loop Pre-Analysis and Annotation Tools}\label{pre_analysis}
We performed a pre-analysis on the loops to obtain tempo, key and genre suggestions. To obtain the tempo information, we followed the same approach of \cite{fontTempo}, parsing the title, description and the tags of the loop for tempo information provided by users.
To propose an initial key and mode to the annotators, we analysed the loops using the algorithm proposed by Faraldo et al. \cite{angelEDM}, which is implemented in the Essentia audio analysis library \cite{essentia}. Finally, by taking the genre information from the textual metadata of the loops, we were able to map some of the sounds to the genres to annotate. The checkboxes were selected for the genres which either were mentioned or had a sub-genre mentioned in the textual metadata. Our annotators were familiar with the annotation procedure and took the pre-annotations only as suggestions to speed-up the annotation process.

In the annotation tool, at the top of the display is the loop's metadata: its unique sound id, title, author's username, and the tags and textual description provided by the author \jordan{(see Fig. \ref{img:annotationtool})}. The waveform of the loop and a playhead is also shown, which are linked to an audio player. The audio player always restarts the playback of the loop when it finishes, and triggers a metronome with the BPM provided in the BPM annotation field. We provide stop, play and pause controls for the loop and metronome and volume controls for the loop. A button which restarts only the metronome is also present. To ease finding a key and mode which suits the loop, we present a synthesizer which plays the chord present in the tonal annotation section. In case the mode selected is ``None'' or ``Unknown'', the synthesizer will just play the root note of the key selected. Using the computer's keyboard, the annotator can cycle through the options for key and mode, in several octaves. Finally, buttons are provided \jordan{for submitting the annotation when it is finished, saving the sound for later and discarding the sound in case it is not a loop.}



\subsection{Dataset Availability}
The loops and corresponding annotations (provided in a JSON file) are publicly available on Zenodo.\footnote{\url{https://zenodo.org/record/3967852}} This dataset can be divided into three subsets, defined by their level of annotations. These are:
\begin{compactitem}
    \item Multiple-annotations (MA): the loops annotated by at least two researchers. It contains 1,472 loops.
    \item Single-annotation (SA): the loops annotated by a single researcher. Currently contains 1,464 loops.
    \item Automatic-annotations (AA): the loops annotated by the analysis algorithms mentioned in Section \ref{pre_analysis}. Contains 9,455 loops\jordan{: the loops in MA and SA and 6,519 more}.
\end{compactitem}
In addition to the main dataset, we provide a repository\footnote{\url{https://github.com/aframires/freesound-loop-annotator}} with the code used for the annotation tool interface and server, the pre-analysis that generated the subset of automatic annotations, and the analysis and potential applications presented in Sections \ref{sec:datasetanalysis}, \ref{pot_app1} and \ref{pot_app2}. 
\section{Dataset Analysis}\label{sec:datasetanalysis}
To understand the diversity and reliability of the dataset, we investigate the distribution of annotated characteristics and inter-annotator agreement.
\subsection{Annotation Distribution}

\begin{figure}[h]
    \centering
    
  \includegraphics[width=0.4\textwidth]{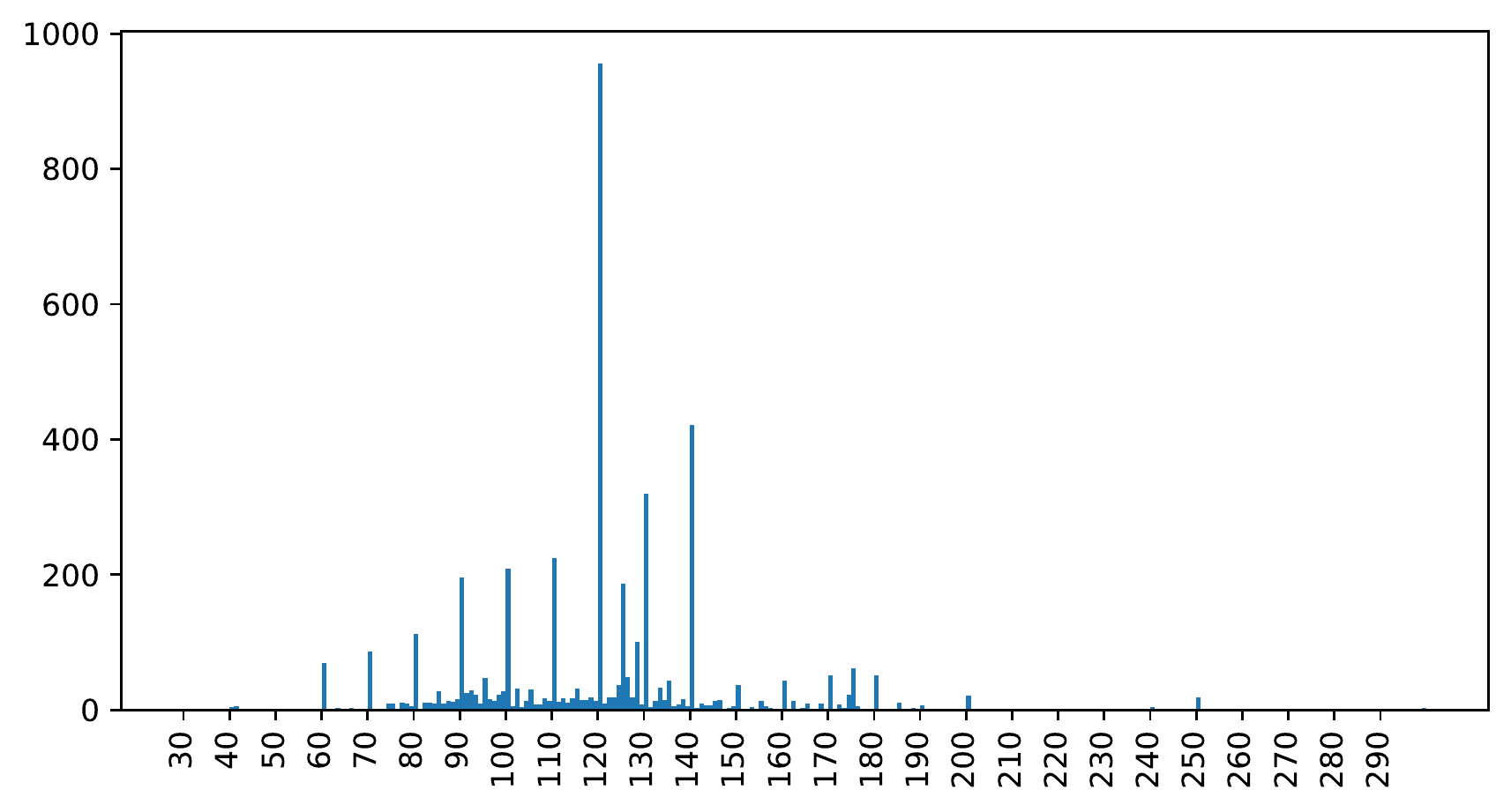}
  \caption{Distribution of BPMs in FSLD.}
  \label{img:histogram}
\end{figure}

The human-annotated part of the dataset contains 1,579 sounds which, in total, have been annotated 2,809 times. The distribution of genres, instrumentation, and keys are shown in Tables \ref{tab:instr} and \ref{tab:keys_dist} and the tempo histogram in Figure \ref{img:histogram}. It is well-balanced in terms of instrument and genre; reasonably balanced in terms of tempo, although 120 bpm dominates;
\jordan{and highly imbalanced in terms of key, with C Major and Minor dominating.}


\begin{table}[h]
\small
    \begin{tabular}{l|c}
        Percussion & 54.95\% \\
        Bass & 19.10\% \\ 
        Chords & 11.90\% \\
        Melody & 21.31\% \\
        FX & 24.80\% \\
        Vocal & 2.29\% \\
    \end{tabular}
    \begin{tabular}{l|c}
        Bass Music & 32.04\% \\
        Live Sounds & 21.38\% \\
        Cinematic & 19.95\% \\
        Global & 14.26\% \\
        Hip-hop & 17.29\% \\
        House/Techno & 29.05\% \\
        Other Dance Music & 25.63\% \\
    \end{tabular}
    \caption{Distribution of the instrument roles and genre in our dataset.}
    \label{tab:instr}
\end{table}


\begin{table}[h]
    \centering
    \small
    \begin{tabular}{l|cccc}

\textbf{Key}   & \textbf{Maj} & \textbf{Min} & \textbf{None} & \textbf{Unknown} \\
\hline
C   & 9.63\% & 8.38\% & 3.65\% & 0.95\% \\
C\# & 1.38\% & 2.84\% & 0.60\% & 0.43\% \\
D   & 3.31\% & 5.37\% & 1.85\% & 0.39\% \\
D\# & 1.29\% & 2.49\% & 0.73\% & 0.21\% \\
E   & 2.28\% & 3.65\% & 1.20\% & 0.26\% \\
F   & 4.64\% & 4.43\% & 1.16\% & 0.34\% \\
F\# & 1.12\% & 2.49\% & 1.12\% & 0.13\% \\
G   & 2.66\% & 4.25\% & 1.93\% & 0.43\% \\
G\# & 1.72\% & 2.58\% & 0.90\% & 0.13\% \\
A   & 3.01\% & 5.50\% & 1.68\% & 0.26\% \\
A\# & 1.50\% & 1.89\% & 0.52\% & 0.04\% \\
B   & 0.82\% & 3.01\% & 0.64\% & 0.21\% \\
    \end{tabular}
    \caption{Distribution of the keys in our dataset.}
    \label{tab:keys_dist}
\end{table}

\subsection{Inter-annotator Agreement}
To measure the agreement of the annotators in our dataset, we measure the inter-annotator agreement for the MA annotations subset. To do this, we use two metrics: proportion of overall agreement (Agr.) for all the annotations, and positive and negative agreement (PA and NA) \cite{agree} for binary classification tasks. The proportion of overall agreement reflects the number of cases when both annotators agree on a label, and is calculated by dividing their number by the total number of annotations. This overall metric does not distinguish the agreement in positive and negative cases, so for the binary annotation tasks we also calculated the positive and negative agreement. The formulas for calculating these are given in Eq. \ref{eq:pa}, where the variables represent the annotations by the annotators (e.g., NP = first annotator answered negative, second positive).
\begin{equation} \label{eq:pa}
    PA = \frac{2PP}{2PP + NP + PN} \,, 
\end{equation}
\begin{equation} \label{eq:na}
    NA = \frac{2NN}{2NN + NP + PN}\,,
\end{equation}


\begin{table}[h]
\small
\begin{tabular}{l|l|ccc}
\textbf{Char.}         &    \textbf{Sub-Char. }           & \textbf{Agr.} & \textbf{PA} & \textbf{NA}         \\
               \hline
Inst.           & Percussion     & 85.16\%      & 86.62\%               & 83.35\%               \\
                & Bass           & 76.73\%      & 45.83\%               & 85.19\%               \\
                & Melody         & 82.33\%      & 60.57\%               & 88.61\%               \\
                & Chords         & 87.40\%      & 47.35\%               & 92.84\%               \\
                & FX             & 72.04\%      & 43.61\%               & 81.41\%               \\
                & Vocal          & 98.66\%      & 71.88\%               & 99.31\%               \\
\hline
Tempo           & BPM            & 87.84\%      & NA                    & NA                    \\
                & Signature      & 97.84\%      & NA                    & NA                    \\
                & Well Cut       & 86.88\%      & 92.73\%               & 32.82\%               \\
                \hline
Key             & Root           & 67.56\%      & NA                    & NA                    \\
                & Mode           & 69.80\%      & NA                    & NA                    \\
                \hline
Genre           & Bass Music     & 69.50\%      & 53.26\%               & 77.37\%               \\
                & Live Sounds    & 80.09\%      & 55.28\%               & 87.19\%               \\
                & Cinematic      & 81.66\%      & 57.14\%               & 88.33\%               \\
                & Global         & 82.33\%      & 51.53\%               & 89.19\%               \\
                & Hip-Hop        & 79.05\%      & 31.30\%               & 87.64\%               \\
                & House/Techno   & 69.35\%      & 48.56\%               & 78.17\%               \\
                & Other          & 73.53\%      & 45.64\%               & 82.50\%               
\end{tabular}
\caption{Inter-annotator agreement for the MA subset.}
\label{tab:agreement}
\end{table}

Table \ref{tab:agreement} presents the results for this analysis.
We can see that overall, the values for the agreement are high. Bass, melody and chords have a lower positive agreement value, despite the high negative agreement. This might indicate that annotators are not able to easily distinguish if an element should fit in one of the 3 roles, but can say when it is not present. The lower value for root key agreement indicates that several keys are used to describe the same sounds. This fits our annotating indications, where we asked annotators to select a key which sounds good with the loop and therefore, personal taste may arise in this choice. Finally, the positive agreement for genres always has values lower than 65\%, which might be due to how genre might be perceived subjectively between annotators.


\section{Benchmarking MIR Tasks} \label{pot_app1}
To demonstrate the usefulness of this dataset, we use it in several short case studies. To benchmark tempo, we followed the evaluation approach of \cite{fontTempo} and used the \emph{Accuracy 1} and \emph{Accuracy 2} presented in \cite{gouyon2006experimental}, together with the \emph{Accuracy 1e} proposed in \cite{fontTempo}. Due to space constraints, here we only report the mean of the 3 accuracies. Full results can be seen in an accompanying
website.\footnote{\label{note1}\url{https://aframires.github.io/freesound-loop-annotator/}} The algorithms selected for the tempo benchmarking were the following (details for each algorithm can be found in respective papers):

\begin{itemize}
    \setlength\itemsep{-0.2em}
    \item \textbf{Percival \cite{percival2014streamlined}:} We use both the original implementation and the one provided in Essentia.
    \item \textbf{Zapata \cite{zapata2014multi}:} Implementation provided in Essentia.
    \item \textbf{Degara \cite{degara2011reliability}:} We also use Essentia's implementation.
    \item \textbf{B\"{o}ck \cite{bock2015}:}We use the 3 variants available in the Madmom library\footnote{\url{https://github.com/CPJKU/madmom}}: COMB, ACF and DBN.
\end{itemize}

We validate tempo estimation algorithms on the 3 proposed subsets. Key estimation is only validated on the MA and SA subsets as we do not have original uploader annotations for key. The MA subset, which has at least 2 annotations per loop, was analysed in two ways: BOTH and EITHER. In BOTH, we run the MIR algorithms exclusively on the loops which have the same labels from both annotators. In EITHER, the output of the algorithm was deemed correct if it was at least one of the annotated labels. The results are presented in Table \ref{tab:tempo}


\begin{table}[h]
\small
\begin{tabular}{l|cccc} 
\textbf{Algorithm}   & \textbf{AA} & \textbf{SA} & \textbf{BOTH} & \textbf{EITHER} \\
\hline
Percival14  & 58.09 & 62.98 & 65.75 & 84.13  \\
Percival14e & 57.82 & 64.00 & 65.49 & 84.98  \\ 
Zapata14    & 51.81 & 58.79 & 58.99 & 77.97  \\
Degara12    & 52.32 & 58.77 & 59.31 & 79.16  \\
Bock15COMB  & 44.42 & 51.17 & 52.92 & 71.35  \\ 
Bock15ACF   & 48.65 & 51.96 & 54.75 & 74.90  \\
Bock15DBN   & 45.76 & 50.60 & 52.32 & 70.90 
\end{tabular}
\caption{Evaluation of tempo estimation algorithms in the proposed subsets.}
\label{tab:tempo}
\end{table}

We can see that the results are similar to the ones obtained in \cite{fontTempo}, with Percival14 having better accuracy across all the datasets. We can see that the accuracy increases from AA to SA, and from SA to BOTH. This might be due to the user-annotated loops having
\jordan{incorrect} annotations; \jordan{it may also be that} when both annotators agree on a tempo,
the tempo is strong and defined. The EITHER evaluation gives the largest accuracies, which may be due to its broader criteria for considering tempos correct.

For benchmarking key estimation algorithms, we used the evaluation metrics from MIREX,\footnote{\url{https://www.music-ir.org/mirex/wiki/2019:Audio_Key_Detection}} which evaluates
\jordan{how \emph{close} the estimated key and the annotated key are to provide an accuracy.}
The algorithms \jordan{compared in}
the evaluation were the following: 
\begin{compactitem}
    \item \textbf{EDMKey \cite{angelEDM}:} We use the implementation in Essentia, with 4 key profiles: Krumhansl\cite{Krumhans}, Temperley\cite{temperley1999}, Shaath\cite{sha2011} and the one proposed in \cite{angelEDM}.
    \item \textbf{EssentiaBasic \cite{essentia}:} Essentia's implementation of the algorithm presented by Gomez \cite{gomez2006tonal}.
    \item \textbf{QMUL \cite{noland2007signal}:} We use the Key Detection implementation available in QM Vamp Plugins.\footnote{\url{https://vamp-plugins.org/plugin-doc/qm-vamp-plugins.html#qm-keydetector}}
\end{compactitem}
In Table \ref{tab:key}, we present part of the results of the key estimation evaluation. Due to lack of space, only the final MIREX scores for each dataset are presented. The full results can be seen in the accompanying
website.

\begin{table}[h]
\small
    \begin{tabular}{l|ccc}
    \textbf{Algorithm}       & \textbf{SA}  & \textbf{BOTH}  & \textbf{EITHER} \\
    \hline
    Edmkey          & 72.26 & 88.25 & 85.63  \\
    EdmkeyKrumhansl & 66.99 & 84.85 & 82.98  \\
    EdmkeyTemperley & 61.46 & 71.78 & 71.77  \\
    EdmkeyShaath    & 72.38 & 88.25 & 85.63  \\
    EssentiaBasic   & 71.25 & 88.80 & 85.30  \\
    QMULKeyDetector & 35.09 & 42.15 & 46.25  \\
    \end{tabular}
    \caption{Evaluation of key estimation algorithms in the proposed subsets.}
    \label{tab:key}
\end{table}

We see that EssentiaBasic and EDMkey are the best performing algorithms here. EDMKey has been specially tuned to be used for EDM, which might make it more suitable to the loops we are annotating. We again see that the accuracy increases from SA to BOTH, which might indicate again that when the key is clear and defined, the algorithms are also able to correctly identify it.

\section{Music Generation and Decomposition}\label{pot_app2}

Another way the dataset is valuable is for creating synthetic datasets of songs
for evaluating loop-extraction algorithms, such as~\cite{loopsep}.
We created 100 random songs, each using 5 random drum loops and 5 non-drum loops (chosen from a subset
of 4/4, 120-bpm, 1-bar, single-instrument loops for which there was no disagreement among the annotators on the instrument role).
Each song is a random arrangement of the loops, either in a \textit{sparse} arrangement, in which one drum and one non-drum loop occurs per bar, or a \textit{dense} one, in which 4 loops occur per bar (i.e., 2 drum and 2 non-drum). For comparison, we also recreated the ``composed'' and ``factorial'' layouts from~\cite{loopsep}. Examples of each layout are shown on the accompanying website.


We used the public implementation\footnote{\url{https://github.com/jblsmith/loopextractor}} of~\cite{loopsep} to extract loops for each song, informed with the true number of loop segments (4 or 10) and the true downbeat boundaries.
The metrics SDR, SIR and SAR
(the signal to distortion, interference and artefacts ratios~\cite{raffel2014mireval})
are reported in
the left part of Table ~\ref{tab:sourcesep}.

These are normally computed by trying all permutations of estimated sources to true sources and using that which maximises the score. This is infeasible for permutations of 10 items, so we first find the permutation that maximises the similarity between the source and true loop spectra.

\begin{table}[h]
\small
    \begin{tabular}{l|ccc|cc}
        \textbf{Layout}     & \textbf{SDR}   &  \textbf{SIR}   & \textbf{SAR}    & \textbf{F1}   & \textbf{Acc.}   \\
        \hline
        Sparse & --5.2  & --3.9   & 15.8   & 0.194  & 0.691      \\
        Dense  & --7.9  & --7.3    & 14.4  & 0.294  & 0.542      \\
        Composed   & 12.5  & 18.4  & 22.6    & 0.585  & 0.546   \\
        Factorial  & 19.8  & 29.2  & 24.1    & 0.560  & 0.551 \\
    \end{tabular}
    \caption{Evaluation of loop source quality (SDR, SIR, SAR) and estimated layouts (F-measure and accuracy) for each song layout.}
    \label{tab:sourcesep}
\end{table}


This permutation is also used to evaluate the quality of the estimated layout. We binarize each row of the estimated layout, using the row's mean as threshold. We then compute the raw accuracy (as in~\cite{loopsep}), but here we propose also using the F-measure, so as not to weight true negatives unduly. The results are in the right columns of Table~\ref{tab:sourcesep}.


SDR, SIR and SAR are all lower for the random 10-part songs than for the 4-part songs, showing that we have created a more challenging testing ground for loop extraction systems. For the layout evaluation, our evaluation makes clear that the raw accuracy gives undue weight to true negatives: the highest accuracy was obtained for the sparse layouts, despite having the lowest F-measure.
This short evaluation is a proof of concept; with more space, we could study the impact of the instrumentation, number of loops, loop duration, and other factors on the separation quality. We can also generate layouts with loops of many durations and evaluate hierarchical loop extraction systems.



\section{Conclusion}\label{sec:conclusion}
In this paper, we presented our work on addressing the lack of standard loop datasets to carry MIR tasks. We presented FSLD, a dataset of audio loops annotated at a level similar to commercial loop collections. These loops are licensed for redistribution and can be used and redistributed for research purposes. We provide a detailed analysis of the dataset and its annotations and provided several use cases for tempo and key benchmarking, music generation and loop separation.
Furthermore, we present the online annotation tool used to build the dataset, and we make it available online so other researchers and the general public can contribute and extend the dataset. 

\section{Acknowledgement}
This research was funded in part by European Union’s Horizon 2020 research and innovation programme under the Marie Skłodowska-Curie grant agreement No765068, MIP-Frontiers and by a grant from the Ministry of Science and Technology, Taiwan (MOST107-2221-E-001-013-MY2).

\bibliography{ISMIRtemplate}

\begin{thebibliography}{10}

\bibitem{analysisRhythmLoops}
Juan~P. {Bello}, Emmanuel {Ravelli}, and Mark~B. {Sandler}.
\newblock Drum sound analysis for the manipulation of rhythm in drum loops.
\newblock In {\em IEEE International Conference on Acoustics Speech and Signal
  Processing Proceedings}, 2006.

\bibitem{bock2015}
Sebastian B{\"o}ck, Florian Krebs, and Gerhard Widmer.
\newblock Accurate tempo estimation based on recurrent neural networks and
  resonating comb filters.
\newblock In {\em Proceedings of the 16th International Society for Music
  Information Retrieval Conference}, 2015.

\bibitem{essentia}
Dmitry Bogdanov, Nicolas Wack, Emilia G{\'o}mez, Sankalp Gulati, Perfecto
  Herrera, Oscar Mayor, Gerard Roma, Justin Salamon, Jose~R. Zapata, and Xavier
  Serra.
\newblock Essentia: an audio analysis library for music information retrieval.
\newblock In {\em Proceedings of the 14th International Society for Music
  Information Retrieval Conference}, 2013.

\bibitem{butler2006unlocking}
Mark~Jonathan Butler.
\newblock {\em Unlocking the groove: Rhythm, meter, and musical design in
  electronic dance music}.
\newblock Indiana University Press, 2006.

\bibitem{chen20ismir}
Bo-Yu Chen, Jordan Smith, and Yi-Hsuan Yang.
\newblock Neural loop combiner: Neural network models for assessing the
  compatibility of loops.
\newblock In {\em Proceedings of the 21st International Society for Music
  Information Retrieval Conference}, 2020.

\bibitem{degara2011reliability}
Norberto Degara, Enrique~Argones R{\'u}a, Antonio Pena, Soledad
  Torres-Guijarro, Matthew~EP Davies, and Mark~D Plumbley.
\newblock Reliability-informed beat tracking of musical signals.
\newblock {\em IEEE Transactions on Audio, Speech, and Language Processing},
  20(1):290--301, 2011.

\bibitem{angelEDM}
{\'A}ngel Faraldo, Emilia G{\'o}mez, Sergi Jord{\`a}, and Perfecto Herrera.
\newblock Key estimation in electronic dance music.
\newblock In {\em 38th European Conference on Information Retrieval}, pages
  335--347. Springer-Verlag, 2016.

\bibitem{agree}
Alvan~R Feinstein and Domenic~V Cicchetti.
\newblock High agreement but low kappa: I. the problems of two paradoxes.
\newblock {\em Journal of clinical epidemiology}, 43(6):543--549, 1990.

\bibitem{FS}
Frederic Font, Gerard Roma, and Xavier Serra.
\newblock Freesound technical demo.
\newblock In {\em ACM International Conference on Multimedia}, 2013.

\bibitem{fontTempo}
Frederic Font and Xavier Serra.
\newblock Tempo estimation for music loops and a simple confidence measure.
\newblock In {\em Proceedings of the 17th International Society for Music
  Information Retrieval Conference}, 2016.

\bibitem{automaticTranscription}
Olivier {Gillet} and Ga\"{e}l {Richard}.
\newblock Automatic transcription of drum loops.
\newblock In {\em IEEE International Conference on Acoustics, Speech, and
  Signal Processing}, 2004.

\bibitem{drumRetrievalSpokenQuery}
Olivier Gillet and Ga\"{e}l Richard.
\newblock Drum loops retrieval from spoken queries.
\newblock {\em Journal of Intelligent Information Systems}, 24(2):159–177,
  2005.

\bibitem{gomez2006tonal}
Emilia G{\'o}mez.
\newblock Tonal description of polyphonic audio for music content processing.
\newblock {\em INFORMS Journal on Computing}, 18(3):294--304, 2006.

\bibitem{gomezMarinPadSad}
Daniel G{\'o}mez-Mar{\'\i}n, Sergi Jord{\`a}, and Perfecto Herrera.
\newblock Pad and sad: Two awareness-weighted rhythmic similarity distances.
\newblock In {\em Proceedings of the 16th International Society for Music
  Information Retrieval Conference}, 2015.

\bibitem{experimentalComparison}
Fabien {Gouyon}, Anssi {Klapuri}, Simon {Dixon}, Miguel {Alonso}, George
  {Tzanetakis}, Christian {Uhle}, and Pedro {Cano}.
\newblock An experimental comparison of audio tempo induction algorithms.
\newblock {\em IEEE Transactions on Audio, Speech, and Language Processing},
  14(5):1832--1844, 2006.

\bibitem{gouyon2006experimental}
Fabien Gouyon, Anssi Klapuri, Simon Dixon, Miguel Alonso, George Tzanetakis,
  Christian Uhle, and Pedro Cano.
\newblock An experimental comparison of audio tempo induction algorithms.
\newblock {\em IEEE Transactions on Audio, Speech, and Language Processing},
  14(5):1832--1844, 2006.

\bibitem{Krumhans}
Carol~L. Krumhans.
\newblock Cognitive foundations of musical pitchl.
\newblock {\em Music Perception: An Interdisciplinary Journal}, 9:476--492, 07
  1992.

\bibitem{lopez-serrano2016}
Patricio L{\'o}pez-Serrano, Christian Dittmar, Jonathan Driedger, and Meinard
  M{\"u}ller.
\newblock Towards modeling and decomposing loop-based electronic music.
\newblock In {\em Proceedings of the 17th International Society for Music
  Information Retrieval Conference}, 2016.

\bibitem{lopez-serrano2017}
Patricio L{\'o}pez-Serrano, Christian Dittmar, and Meinard M{\"u}ller.
\newblock Finding drum breaks in digital music recordings.
\newblock In {\em Proceedings of the International Symposium on Computer Music
  Multidisciplinary Research}, 2017.

\bibitem{noland2007signal}
Katy Noland and Mark Sandler.
\newblock Signal processing parameters for tonality estimation.
\newblock In {\em Audio Engineering Society Convention 122}, May 2007.

\bibitem{percival2014streamlined}
Graham Percival and George Tzanetakis.
\newblock Streamlined tempo estimation based on autocorrelation and
  cross-correlation with pulses.
\newblock {\em IEEE/ACM Transactions on Audio, Speech, and Language
  Processing}, 22(12):1765--1776, 2014.

\bibitem{raffel2014mireval}
Colin Raffel, Brian McFee, Eric~J. Humphrey, Justin Salamon, Oriol Nieto, Dawen
  Liang, and Daniel~PW Ellis.
\newblock {mir\_eval}: {A} transparent implementation of common {MIR} metrics.
\newblock In {\em Proceedings of the 15th International Society for Music
  Information Retrieval Conference}, 2014.

\bibitem{modificationRhythmLoops}
Emmanuel {Ravelli}, Juan~P. {Bello}, and Mark {Sandler}.
\newblock Automatic rhythm modification of drum loops.
\newblock {\em IEEE Signal Processing Letters}, 14(4):228--231, 2007.

\bibitem{roma_gerard_2015_3674147}
Gerard Roma.
\newblock {\em {Algorithms and representations for supporting online music
  creation with large-scale audio databases}}.
\newblock PhD thesis, Universitat Pompeu Fabra, June 2015.

\bibitem{sha2011}
ibrahim Sha’ath.
\newblock Estimation of key in digital music recordings.
\newblock Master's thesis, Birkbeck College, University of London, 2011.

\bibitem{18chiloopmaker}
Z.~Shi and G.~J. Mysore.
\newblock {LoopMaker}: Automatic creation of music loops from pre-recorded
  music.
\newblock In {\em Proceedings of SIGCHI International Conference on Human
  Factors in Computing Systems}, 2018.

\bibitem{loopsep}
Jordan B.~L. {Smith} and Masataka {Goto}.
\newblock Nonnegative tensor factorization for source separation of loops in
  audio.
\newblock In {\em 2018 IEEE International Conference on Acoustics, Speech and
  Signal Processing}, pages 171--175, 2018.

\bibitem{LoopsasGenreResources}
Glenn Stillar.
\newblock Loops as genre resources.
\newblock {\em Folia Linguistica}, 39(1-2):197 -- 212, 2005.

\bibitem{08icme_loop_explore}
S.~Streich and B.~S. Ong.
\newblock A music loop explorer system.
\newblock In {\em Proceedings of International Computer Music Conference},
  2008.

\bibitem{temperley1999}
David Temperley.
\newblock What's key for key? the {Krumhansl-Schmuckler} key-finding algorithm
  reconsidered.
\newblock {\em Music Perception: An Interdisciplinary Journal}, 17(1):65--100,
  1999.

\bibitem{zapata2014multi}
Jos{\'e}~R Zapata, Matthew~EP Davies, and Emilia G{\'o}mez.
\newblock Multi-feature beat tracking.
\newblock {\em IEEE/ACM Transactions on Audio, Speech, and Language
  Processing}, 22(4):816--825, 2014.

\end{thebibliography}

%
%
%
%
\end{document}